\documentclass{iaus}
\usepackage{graphicx}

\title[Photometric observations of TrES-3b] 
{New photometric observations of the transiting extrasolar planet TrES-3b}

\author[M. Va\v{n}ko et~al.]   
{M.\,Va\v{n}ko$^1$,
M.\,Jakub\'ik$^1$,
T.\,Krej\v{c}ov\'a$^2$,
        G.\,Maciejewski$^3$,
        J.\,Budaj$^1$,
        T.\,Pribulla$^1$,
        J.\,Ohlert$^{4,5}$,
        St.\,Raetz$^6$,
        V.\,Krushevska$^7$,
        P.\,Dubovsky$^8$}

\affiliation
{$^1$Astronomical Institute of the Slovak Academy of Sciences,
Slovakia, email: {\tt vanko@ta3.sk} \\[\affilskip]
$^2$Masaryk University, Department of Theoretical Physics and Astrophysics,
602 00 Brno, Czech Republic \\[\affilskip]
$^3$Toru\'n Centre for Astronomy, N. Copernicus University Gagarina 11,
87100, Toru\'n, Poland \\[\affilskip]
$^4$University of Applied Sciences, Wilhelm-Leuschner-Strasse 13, 61169 Friedberg, Germany\\[\affilskip]
$^5$Michael Adrian Observatory, Astronomie Stiftung Trebur, Fichtenstrasse 7, 65468 Trebur, Germany\\[\affilskip]
$^6$Astrophysikalisches Institut und Universit\"ats-Sternwarte, Schillerg\"sschen 2-3, 07745 Jena, Germany\\[\affilskip]
$^7$Main Astronomical Observatory of National Academy of Sciences of Ukraine, 27 Akademika Zabolotnoho St. 03680 Kyiv,
Ukraine\\[\affilskip]
$^8$Vihorlat Observatory, Mierov\'a 4, Humenn\'e, Slovakia\\[\affilskip]}

\pubyear{2011}
\volume{xxx}  
\pagerange{000 --000}
\jname{Title of your IAU Symposium}
\editors{A.C. Editor, B.D. Editor \& C.E. Editor, eds.}
\begin{document}

\maketitle

\begin{abstract}
We present new transit observations of the transiting exoplanet TrES-3b
obtained in the range 2009 -- 2011 at several observatories.
The orbital parameters of the system were redetermined and the new linear
ephemeris was calculated. We performed numerical simulations for
studying the long-term stability of orbits. 
\keywords{exoplanets, fundamental parameters, individual (TrES-3b)}
\end{abstract}

\firstsection 

\section{Introduction}

TrES-3b is one of the more massive transiting extrasolar planets. 
The planetary system consists of a nearby G-type dwarf and a massive 
hot Jupiter with an orbital period of 1.3 days.
It was discovered by \cite[O'Donovan et al. (2007)]{O'Donovan07} 
and a discovery-quality light curve has also been obtained by the 
SuperWASP survey (\cite[Collier Cameron et al. 2007]{CollierCameron07}). 
Follow-up transit photometry has been presented by 
\cite[Sozzetti et al. (2009)]{Sozzetti09} and \cite[Gibson et al. (2009)]{Gibson09}. 

\section{Observations \& Results}

All observations used in this study were carried out at the several
observatories: Star\'a Lesn\'a (Slovakia), Toru\'n Center for Astronomy (Poland),
Michael Adrian Observatory (Germany), University Observatory Jena (Germany)
and Vihorlat Observatory (Slovakia). We used telescopes with diameters of the 
primary mirrors in the range of 30--120~cm and optical CCD-cameras (RI-bands).
In order to have a homogeneous dataset we have used the composition of 560 data points 
from two nights obtained at one observatory in the same filter (see Fig.~1). 
To obtain an analytical transit LC we assumed the quadratic limb darkening
law. The limb darkening coefficients $c_{1}$ and $c_{2}$ were linearly interpolated from 
\cite[Claret (2000)]{Claret00} for the following stellar parameters: $T_{\rm eff}$~=~5650~K, 
$log$~(g)~=~4.4 and [Fe/H]~=~-0.19 (based on the results of Sozzetti et al. 2009).
Finally, the orbital parameters of the system were redetermined: 
$R_p/R_*$~=~0.1819(20), $R_*/a$~=~0.1785(70), $i$~=~80.9(7), 
$R_*$~=~0.876$^{+0.008}_{-0.016}$ $R_{\odot}$ and 
$R_p$~=~1.551$^{+0.014}_{-0.028}$ $R_{J}$. To estimate the uncertainties
of parameters we have used the Monte Carlo simulation method. Based on the 
transits obtained at other observatories we calculated following new linear 
ephemeris for TrES-3b: $T_c(E)$~=~2454538.5807(1) + E $\times$ 1.30618595(15).

\begin{figure}[!t]
\begin{minipage}[htb]{0.47\linewidth}
\centering
\includegraphics[width=6cm]{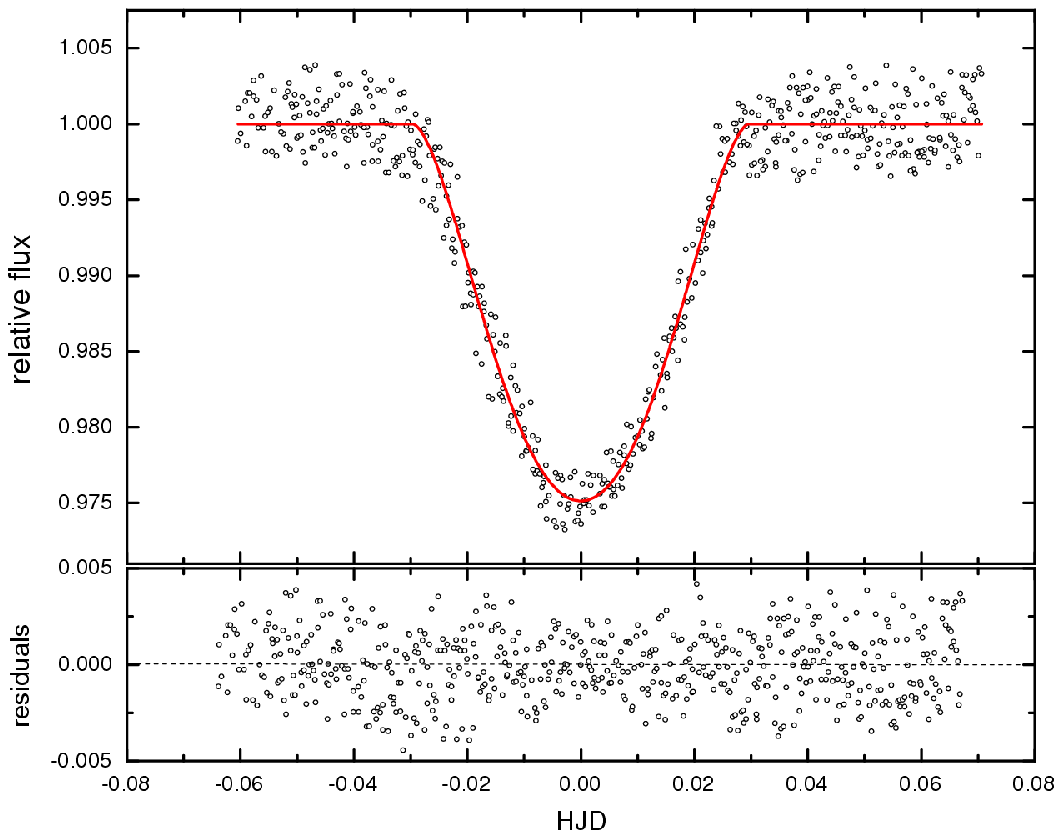}
\end{minipage}
\begin{minipage}[htb]{0.47\linewidth}
\centering
\includegraphics[height=7cm,angle=-90]{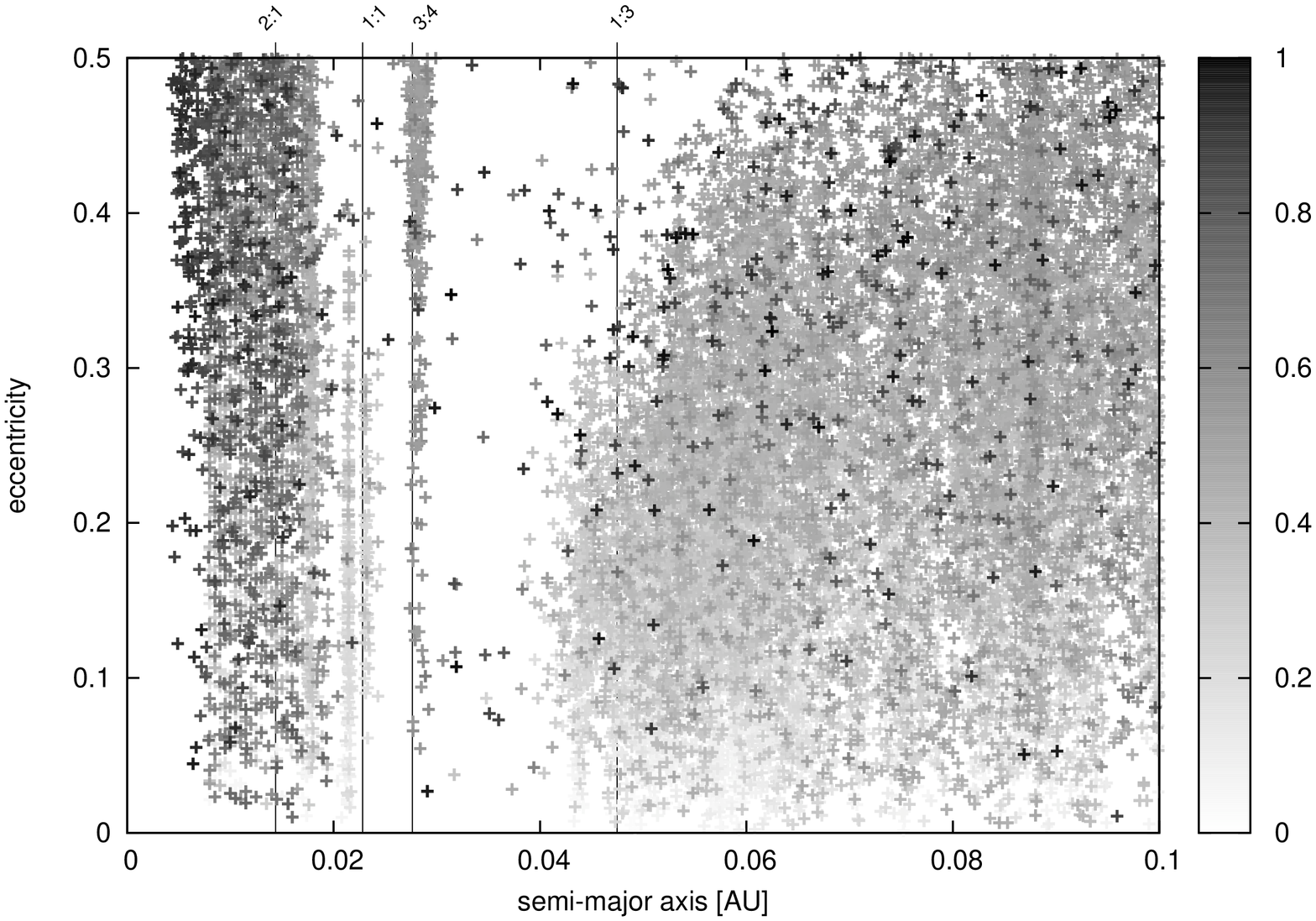}
\end{minipage}
\caption{
{\bf Left} Composition of 560 data points from the two nights: 13/07/10 and
20/08/10 (dd/mm/yy) obtained at Michael Adrian Observatory  in Trebur (Germany). 
{\bf Right} Stability plot in the $a-e$ plane showing the maximum eccentricity. 
}

\label{f1}
\end{figure} 

\noindent 
In this part of our work, we investigate the gravitational influence of
TrES-3b on a potential second planet in the system. \cite[Gibson et al.
(2009)]{Gibson09} claimed, that their data are sensitive enough to probe 
small mass planets on circular orbits near the 2:1 mean-motion resonance 
(MMR) with TrES-3b. We performed numerical simulations for studying the stability 
of orbits and checking their chaotic behavior using the method of maximum eccentricity 
(e.g. \cite[Dvorak et al. 2003]{Dvorak_2003}). We have generated $10^5$ 
massless particles to represent small planets (Earth-like) in this system.
The long-term stability plot in $a-e$ plane showing the maximum eccentricity
after 500 years (about 140 000 revolutions of TrES-3b around the parent star) 
is shown on Fig.\,\ref{f1} (right plot). We found that the region from 
$0.02\,$AU to $0.04\,$AU is almost completely depleted, 
excepted the regions near 1:1 and 3:4 MMRs. The region near 2:1 MMR is richly populated, 
but the particles have relatively high eccentricities and inclinations, 
thus the stability and also the detection using
the TTV method is questionable. More stable regions are beyond the 1:3 MMR, where the 
gravitational influence of the TrES-3b is weak. More detailed study
of the long-term stability in this system (especially near 2:1 MMR) will be 
presented in the future work.\\ 


\noindent \textbf{Acknowledgements } 
This work has been supported by grant, VEGA No. 2/0078/10, 2/0074/09, 2/0094/11. 
PD received the support from APVV grant LPP-0049-06 and
LPP-0024-09. TK thanks the grant GA \v{C}R GD205/08/H005. 
GM acknowledges Iuventus Plus grant IP2010 023070.
SR thanks the German National Science Foundation (DFG) for support in project  NE 515 / 33-1.


\begin{thebibliography}{}

\bibitem[Collier Cameron et al. (2007)]{CollierCameron07}
{Collier Cameron, A., Wilson, D.M., West, R.G., Hebb, L., et al.} 2007,
\textit{MNRAS}, 380, 1230

\bibitem[Claret (2000)]{Claret00}
{Claret, A.} 2000, 
\textit{A\&A}, 363, 1081

\bibitem[Dvorak \etal\ 2003]{Dvorak_2003}
{Dvorak, R., Pilat-Lohinger, E., Funk, B., \& Freistetter, F.} 2003,
\textit{A\&A}, 398, L1

\bibitem[Gibson et al. (2009)]{Gibson09}
{Gibson, N.P., Pollacco, D., Simpson, E.K., Barros, S., et al.} 2009,
\textit{ApJ}, 700, 1078

\bibitem[O'Donovan et al. (2007)]{O'Donovan07}
{O'Donovan, T.F., Charbonneau, D., Bakos, G., Mandushev, G., et al.} 2007,
\textit{ApJ}, 663, L37

\bibitem[Sozzetti et al. (2009)]{Sozzetti09}
{Sozzetti, A., Torres, G., Charbonneau, D., Winn, J.N., et al.} 2009,
\textit{ApJ}, 691, 1145

\end{thebibliography}
\end{document}